\documentclass[12pt]{elsarticle}
\usepackage{amsmath,amsfonts,amssymb,graphicx,hyperref,textcomp,color}

\usepackage{amsfonts,hyperref,textcomp,color,multirow,adjustbox}
\usepackage[bottom]{footmisc}

\usepackage{graphicx}
\usepackage{dcolumn}
\usepackage{bm}
\usepackage{epsfig}
\usepackage{amsmath}
\usepackage{pgf}
\usepackage{float}
\usepackage{amssymb}
\usepackage{subcaption}
\usepackage{tikz}
\usepackage{mathptmx}

\renewcommand{\epsilon}{\varepsilon}

\begin{document}

\begin{frontmatter}

\title{Information theory and player archetype choice in Hearthstone}

\author[rvt]{Mathew Zuparic}
\author[rvr]{Duy Khuu}
\author[rvs]{Tzachi Zach}
\address[rvt]{Defence Science and Technology Group, Canberra, ACT 2600, Australia}
\address[rvr]{Australian National University, Canberra ACT 2601 Australia}
\address[rvs]{Ohio State University, Columbus OH 43210 United States}

\begin{abstract}
Using three years of game data of the online collectible card game \textit{Hearthstone}, we analyse the evolution of the game's system over the period 2016--2019. By considering the frequencies that archetypes are played, and their corresponding win-rates, we are able to provide narratives of the system-wide changes that have occurred over time, and player reactions to them. Applying the archetype frequencies to analyse the system's Shannon entropy, we characterise the salient features of the time series of player choice. Paying particular attention to how entropy is affected during periods of both small and large-scale change, we are able to demonstrate the effects of increased player experimentation before popular decks and tactics emerge. Furthermore, constructing conditional probabilities that simulate understandable player behaviour, we analyse the system's information storage and test the explain-ability of current player choice based on previous decision-making.
\end{abstract}



\end{frontmatter}



\section{Introduction}
\label{intro}

Player choice in most adversarial games (\textit{Chess}, \textit{Go}, \textit{etc.}) is limited to on-board actions, where the distinguishing factor of success is typically experience enabling players to make better choices over the course of a match. Collectible card games (CCGs) such as \textit{Magic: the Gathering}, \textit{Yu-Gi-Oh!} and \textit{Hearthstone} require players construct their specific deck of cards \textit{before} they engage other players. Decks consist of a limited number of cards (approximately thirty) from a potential pool of thousands. The act of constructing a deck is arguably the most meaningful choice a CCG player will make, usually determining the tactics that will be pursued during a match. For CCGs with a significant number of players, popular decks (also referred to as \textit{archetypes}) inevitably emerge over time.

\begin{figure}
    \centering
    \includegraphics[width=8.5cm]{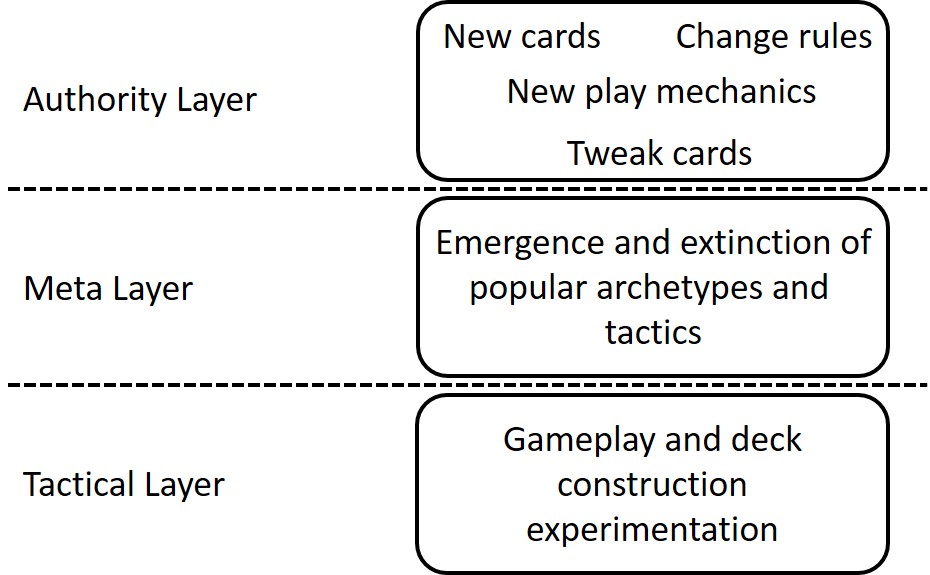}
    \caption{Abstraction of the online CCG system.}
    \label{fig:landscape}
\end{figure}

Online CCGs exist within a substantial system. \textit{Hearthstone} for example currently boasts approximately one hundred million players world-wide. Figure \ref{fig:landscape} presents an abstraction of this system, divided into three layers: \textit{tactical}, \textit{meta} and \textit{authority}. Gameplay occurs at the tactical layer, where players attempt to make the best moves given the cards in their hand. Deck construction also occurs at this layer, where players choose cards based on their preferred style of play, relying on past experiences to refine their choice of cards. The meta for an online CCG can be articulated as the emergence of archetypes and corresponding tactics over time. As explained by Carter \textit{et al.} \cite{carter12}, though the meta is \textit{peripheral} to the rules and mechanics of the game itself, it describes the environment that players will experience. The authority layer in Figure \ref{fig:landscape} is the remit of the publisher of the CCG. At various times, the publisher will initiate change into the tactical layer of the CCG, either by changing cards, adding new cards or changing rules and/or play mechanics.

\subsection{Past research}

Due to the underlying difficulty involved in deck construction, the majority of past research in CCGs has been focused on the discovery of optimal decks and/or determining the best gameplay options. Demonstrating this difficulty, in September 2018 standard players of \textit{Hearthstone} could construct approximately $4.11 \times 10^{72}$ unique decks. To efficiently explore this sample space, Garc\'{i}a-S\'{a}nchez \textit{et al.} \cite{Garcia16} applied evolutionary algorithms to continuously evolve initial decks and showed a noticeable improvement in win-rates against the majority of other deck archetypes. In \cite{Garcia18} the authors used evolutionary algorithms to create decks which outperformed archetypes within the meta when played by an artificial intelligence (AI). Most recently the authors in \cite{Garcia20} tested their methods in an international \textit{Hearthstone} AI competition, placing in the top 6\% of entrants. Bursztein \cite{Bursztein16} demonstrated an algorithm which predicted opponent moves in \textit{Hearthstone}, boasting over 95\% accuracy after the second turn. Remarkably, the publisher requested that the author not publicly release the algorithm as it was game breaking. Stiegler \textit{et al.} \cite{Stiegler16} applied a utility system to automatically construct decks which considered metrics relating to gameplay (cost effectiveness, card synergies, \textit{etc.}), as well as popularity of cards in the meta. The authors found that their algorithm was able to complete deck skeletons into currently popular archetypes. Fontaine \textit{et al.} \cite{Fontaine19} applied an algorithm which incentivised both novelty and performance to imitate player decision-making and determine a set of popular decks. Once decks were established the authors negatively affected (\textit{nerfed}) some commonly selected cards to limit their viability. Counter-intuitively, the authors found that the algorithm included some cards \textit{more} frequently after they were nerfed, suggesting that even an objectively negative change may have a positive impact on the perception of a card's viability. Bhatt \textit{et al.} \cite{bhatt18} found that decks constructed by their algorithm possessed some degree of generality, performing well against decks not in the current meta. Focusing on the meta layer of CCGs, de Mesentier Silva \textit{et al.} \cite{silva19} applied an evolutionary algorithm to understand the impacts brought on the meta by improving and/or nerfing various cards. While it was possible to balance the meta after initiating change, the authors noted that too much change was difficult to resolve. 

\subsection{Intent of this work}
This work focuses on the CCG system itself as presented in Figure \ref{fig:landscape} --- to the best of our knowledge a topic not actively studied thus far. \textit{At a system-wide level we seek to understand how changes enacted to the CCG by the publishers influence player decision-making.} We demonstrate how small and large-scale changes initiated in the tactical layer influence the evolution of the CCG's meta, and by extension the decisions players make in deck construction. To enable this understanding we use three years of gameplay data from the CCG \textit{Hearthstone}, considering both the frequencies that deck archetypes are played, and their corresponding win-rates over the 2016--2019 period. To understand the effects that changes enacted by the publishers have on the \textit{Hearthstone} system, we focus on various information-theoretic measures as they are a method to quantify the amount of surprise, randomness and complexity in an entire system. Analysing the \textit{Hearthstone} meta through the lens of information entropy, we understand and characterise the evolution of complexity and uncertainty in the meta at any given time. Furthermore, by considering the information storage exhibited within \textit{Hearthstone's} meta, we are able to estimate how much previous player decision-making explains the underlying structure seen in the current state of the meta.

\subsection{Tactical elements and card properties}
\label{tactical}
Though we focus on \textit{Hearthstone}, many of the elements discussed in this section apply to the majority of CCGs. During a match, two adversaries take turns selecting and \textit{playing} cards from their \textit{hand}. At the start of each turn a specific number of cards are randomly dealt to the player's hand, drawn from the deck which has been constructed by the player beforehand. During a player's turn cards are \textit{activated} by spending a predefined amount of in-game resource (\textit{mana}) which is replenished at the end of the player's turn. Cards that have been activated are then sent to the player's \textit{discard-pile}, out of play for the remainder of the match. The ultimate goal for each player is to reduce their adversary's \textit{health} to zero. 

Cards generally fall under two categories: \textit{minions} and \textit{spells}. Minions give the player controllable characters which enable a range of defence and offence options. Spell cards range from single-use damage-dealing cards, to cards which perform sustained effects over multiple turns. Specific to \textit{Hearthstone} \cite{Bursztein16, silva19} are also \textit{weapon} cards, equippable by the player, \textit{secret} cards, similar to spells but triggered once specific conditions are met, and \textit{hero} cards, changing the properties of the player's in-game avatar. The \textit{synergies} between card properties also affect player choice during deck construction. For instance, the activation of many \textit{high value} cards may require complex conditions, which can only be satisfied if other specific cards are present. For more information regarding \textit{Hearthstone} card properties we refer the reader to \cite{wikiHearth}.

\subsection{Archetypes and character classes}
\label{deckarch}
Archetypes in CCGs \textit{generally} fit into the following three categories:
\begin{itemize}
    \item{\textit{Aggro} archetypes rely on aggressive tactics to achieve victory. They typically focus on low cost cards with the intent of overwhelming the adversary in the early stages of the match. Aggro players that cannot maintain significant tempo in the early-to-mid stages of the match typically lose.} 
    \item{\textit{Control} archetypes rely on relatively high-cost and high-value cards to win in the later stages of a game. The moniker \textit{control} comes from the archetype's strategy in the early-to-mid game of countering a variety of play-styles, thus granting player the time needed to initiate the intended late-game finishing tactics.}
    \item{\textit{Combo} archetypes generally rely on cards which contain synergies, with the intent to knock out the opponent by playing a number of cards in conjunction with each other to generate devastating effects. Much like control, combo archetypes must have some form of counter for aggressive early-game play-styles, but mirroring aggro archetypes, they also rely on knocking out control archetypes before their high-cost high-value cards are activated.}
\end{itemize}

Specific to \textit{Hearthstone} in the 2016--2019 period, players must additionally choose one of nine character classes: \textit{Druid}, \textit{Hunter}, \textit{Mage}, \textit{Paladin}, \textit{Priest}, \textit{Rogue}, \textit{Shaman}, \textit{Warlock} and \textit{Warrior}. Characters provide unique abilities, and grant the player character-specific cards. Some characters generally favour specific archetypes due to these exclusive cards. For instance \textit{Mages} gravitate towards \textit{control} archetypes due to
the considerable range of spell cards resulting in their ability to deal with a large number of adversary minions.

In a recent survey of online CCG players, Turkay and Adinolf \cite{Turkay18} established that player motivations fell under 4 categories: \textit{immersion}, \textit{competition}, \textit{socialisation} and \textit{strategy development}. Thus, players can be motivated by more than simply winning, including finding combinations of cards particularly fun to play. A \textit{Hearthstone} specific example of this is the card \textit{Marin the Fox}, which when summoned creates a treasure chest that, once destroyed, grants the player with one of a number of extraordinarily powerful cards. It was recognised that successful implementation of this tactic posed many risks as a number of archetypes possessed abilities which would allow the opponent to steal the resulting powerful cards for themselves. Nevertheless, despite considerable risks this card did appear in a number of decks due to how satisfying the chest's rewards were if obtained. 

\subsection{Information-theoretic measures}

Information-theoretic measures such as Shannon entropy quantify the amount of \textit{surprise} or \textit{randomness} exhibited in a system \cite{Arora81, MacKay03}. \textit{Hearthstone's} meta is a complex system where archetypes emerge and disappear, and a range of behaviours can be exhibited as time progresses. Measuring the amount of randomness displayed in the meta over time enables appreciation of how balanced active archetypes are, how effective recent changes were, and ultimately help characterise the state of the \textit{Hearthstone} system. Past examples of Shannon entropy offering insights into complex systems include: Miranskyy \textit{et al.} \cite{Miranskyy12} who used entropy measures to understand and compare rare events in defective software; Cao \textit{et al.} \cite{Cao2014} who developed Shannon entropy-based measures on graphs to understand the underlying complexity of graph families; and Aggarwal \cite{Aggarwal2019} who recently applied generalised Shannon entropy measures to provide decision support in the face of multiple criteria that were often conflicting.  

Associated with information entropy is the concept of system \textit{criticality} \cite{hohenberg77}, sometimes referred to as the \textit{edge of chaos} \cite{roli18}. In mathematical \cite{Kallon18}, physical \cite{Polani13}, biological \cite{Mora11} and computational \cite{Langton90} systems, amongst others, criticality refers to the system being able to respond and adapt to a rapidly changing environment. Intuitively, it can be viewed as a dynamical system cycling through periods of relatively low and high entropy, spending the majority of its time in intermediate entropy values. For an introduction to this topic refer to \cite{crutch90, prok09}. 

We additionally apply the concepts of distributed information storage \cite{Mitchell93}, closely related to information transfer \cite{Gencaga15}. The systematic explanation of how information is stored, processed and transferred in distributed systems began in earnest with Schreiber's \cite{Schreiber00} landmark work on \textit{information transfer entropy}, mathematically defining how information is transferred between distinct processes in distributed systems. Information transfer has since been applied to great effect in a wide range of applications, including neuroscience \cite{Vicente11}, multi-agent dynamics \cite{Lizier08}, and social media \cite{Steeg11}. A decade after Schreiber's result, Lizier \textit{et al.} \cite{Lizier2012} introduced \textit{local active information storage} (LAIS) to distinguish the dynamics displayed in cellular automata. LAIS has since been applied to understand how information storage properties affect network structure in biological and artificial networks \cite{Lizier2012b}, and characterise normal and diseased states in cardiovascular and cerebrovascular regulation \cite{Faes2013}. Wu \textit{et al.} \cite{Wu2013} demonstrated the utility of \textit{localising} other information-theoretic measures as a means of overcoming their known weaknesses regarding image recognition.

This work applies LAIS to appreciate how much the past state of the \textit{Hearthstone} meta contributes to its current state.
This is motivated by a number of studies which apply LAIS to explore similarly themed questions on a number of complex systems. These include Wibral \textit{et al.} \cite{Wibral2014} who measured the local time and space voltage neurologically generated by stimulating the visual cortex of an anaesthetised cat. The spatio-temporal structure of the corresponding LAIS data characterised how the onset of visual stimulus led to spatio-temporal surprise (or misinformation) about the proceeding visual outcomes. Wang \textit{et al.} \cite{Wang12} explored collective memory/storage via an information-theoretic characterisation of cascades within the dynamics of simulated swarms. Using the interpretation that the LAIS of a system component characterises the amount of past data used to predict the component's next state, the authors calculated the system-wide \textit{active information storage} (AIS) by taking the expectation value over all component states at any time period. They verified a long-held conjecture that information, used for computation by the swarm, cascaded via waves rippling through the swarm, and found that higher values of storage generally correlate with greater dynamic coordination. Cliff \textit{et al.} \cite{Cliff17} explored the AIS within a multi-agent team by analysing implicit team interactions. The authors noted that when an agent's AIS values were high its movements were largely predictable.

\subsection{Mathematical preliminaries}

For a set of $K+1$ time-ordered states, $\{ X^T, X^{T-1}, \dots, X^{T-K} \}$, the LAIS of the state $X^T$ at time $T$, based on its past $K$ states, is given by
\begin{equation}
    a_K(X^T)= \log_2 \frac{\mathcal{P}(X^T|X^{T-1},\dots,X^{T-K})}{\mathcal{P}(X^T)}.
    \label{LAISdef}
\end{equation}
Positive values of Eq.(\ref{LAISdef}) imply that the past states of the variable provide information and positively correlate with the current state. Conversely, negative values of LAIS indicate that the variable's past history does not correlate with its next state and is synonymous with surprise. The expectation value of the LAIS (the AIS)
\begin{equation}
     A^T_K \equiv \langle  a_K(X^T) \rangle = \left[ \prod^K_{n=1} \sum_{X^{T-n} \in \mathcal{X}^{T-n}} \right] \mathcal{P}(X^T, X^{T-1},\dots,X^{T-K}) a_K(X^T)
\end{equation}
is the \textit{explain-ability} \cite{Lizier2012} of the information in the system. That is, when compared to the corresponding Shannon entropy, the AIS gives the amount of information in the current system that \textit{is} explainable by the results of the previous time step(s). To further clarify this concept of explain-ability, complementary to AIS is the \textit{entropy rate} $\mathcal{E}^T_K$, given by
\begin{equation}
    \mathcal{E}^T_K = -\langle \log_2 \mathcal{P}(X^T|X^{T-1},\dots,X^{T-K}) \rangle.
    \label{LERdef}
\end{equation}
When compared to the corresponding Shannon entropy, the entropy rate gives the amount of information in the current system which \textit{is not} explainable by the results of the previous time step(s).

Following Lizier \textit{et al.} \cite{Lizier2012} and Crutchfield and Feldman \cite{Crutch03}, the contrast between what is explainable and what isn't in the system is made clear by the following duality relation between Shannon entropy --- labeled $ \mathcal{H}(\mathcal{X}^T)$ --- of the current state, AIS, and the entropy rate via
\begin{equation}
 \mathcal{H}(\mathcal{X}^T)=A^T_K + \mathcal{E}^T_K.
\label{duality}
\end{equation}
Thus by Eq.(\ref{duality}) the percentage of information within the system which is explainable by past results is given by $A^T_K / \mathcal{H}(\mathcal{X}^T)$, and the remaining $\mathcal{E}^T_K/ \mathcal{H}(\mathcal{X}^T)$ being the percentage of information not explained by past results.

\subsection{Outline of the paper}
\label{sec:past}
In the next section we detail the \textit{Vicious Syndicate} website which is the source of the \textit{Hearthstone} data considered in this work. Using this data, we then construct sample timelines of some deck archetypes, demonstrating the dynamic evolution of the meta over time. We then look at the data through the lens of Shannon entropy. In Section \ref{sec:AIS} we construct conditional probabilities which simulate relatively simple, but nonetheless understandable, player deck choice behaviour. These conditional probabilities are used to define the system-wide AIS values per time period, ultimately applying Eq.(\ref{duality}) to understand how much of \textit{Hearthstone's} past state of deck frequencies and win-rates contributes to its current state. In Section \ref{conc} we offer further discussion and detail potential future work.

\section{Data explanation and exploration}

\subsection{Data collection and preparation: Vicious Syndicate}
\label{collection}

\textit{Vicious Syndicate} has been collecting \textit{Hearthstone} game data systematically since May 2016. The data was used to produce weekly \textit{Data Reaper Reports} about the state of the \textit{Hearthstone} meta-game \cite{VSDRR}. Breaks in reporting occur near the release of new content by \textit{Hearthstone's} publishers. To contribute game data, players are asked to install a small plugin that records their game play. That data is transmitted to the \textit{Vicious Syndicate} team to be processed. 

During any week between 2000 to 5000 players contributed game data, with tens of thousands of games being processed to produce reports. Specific numbers of contributing players and processed games can be found in each of the corresponding \textit{Data Reaper Reports} \cite{VSDRR}. Only games of rank 15 and above are included for reporting purposes. Only opponent archetypes are included for frequency reporting so as to avoid potential over-representation of archetypes favoured by players who contributed data \cite{VSdata}. Deck identification algorithms are applied to classify archetypes based on the cards played during a match. Though not every game provided a definitive identification, algorithms achieved a high success rate ($> 95\%$) in archetype classification. Additionally,  win-rates were evaluated by taking the average of archetype match-ups from the player perspective (those who contribute data) and the opponent perspective. The data was additionally filtered to exclusively include the archetypes which battled \textit{all other} archetypes present in the meta at least twenty times per reporting period.

\subsection{Archetype timelines}
\label{sec:timeline}

\begin{figure}
    \centering
    \includegraphics[width=16cm]{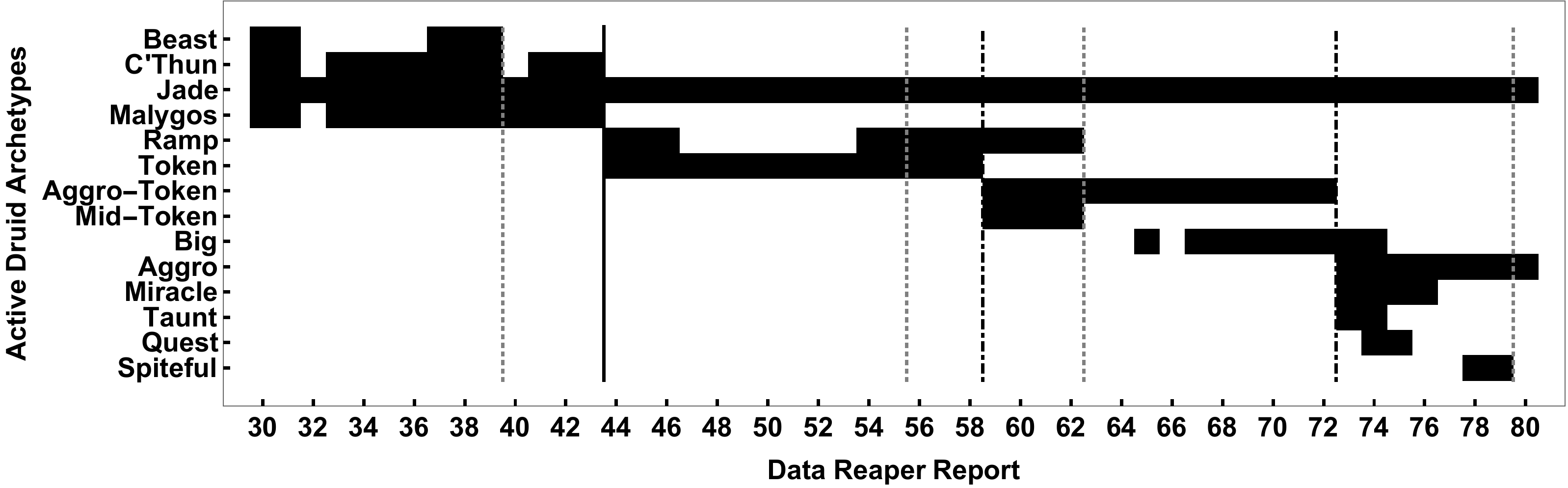}
    \caption{Timeline of the \textit{Druid}-based archetypes present in the meta. Vertical lines represent system changes by the publisher --- \textbf{solid lines} signify release of an expansion (new cards and game mechanics) \textit{in addition to} a rotation of a number of older cards out of the standard mode; \textbf{dot-dashed lines} signify the release of an expansion; and, \textbf{dotted lines} signify release of balance patches (changes to existing cards). The horizontal axis corresponds to the \textit{Data Reaper Report} from which the data is drawn. The data for the first entry, report 30, was collected over the period 14-20 December 2016. The data for the final entry, report 80, was collected over the period 6-13 February 2018.}
    \label{fig:druidTL}
\end{figure}

To illustrate an example Figure \ref{fig:druidTL} depicts the timeline (from December 2016 to February 2018) of the \textit{Druid}-based archetypes present in the meta. Each archetype (14 in total) is noted on the vertical axis as they appear in chronological order. The horizontal axis corresponds to the \textit{Data Reaper Report} from which the data is drawn. Each black horizontal bar designates the appearance of that particular archetype in the meta over the appropriate time period. Each vertical line represents a specific change to the system: \textbf{solid lines} indicate the release of an expansion (with new cards and game mechanics) \textit{in addition to} a rotation of a significant proportion of older cards out of the standard mode; \textbf{dot-dashed lines} signify the release of an expansion; and \textbf{dotted lines} signify release of balance patches (changes to a number of existing cards).  

The change that occurred after report $43$ in Figure \ref{fig:druidTL} was due to the release of the \textit{Journey to Un'Goro} expansion which introduced 135 new cards (some with new play mechanics) to the game. Additionally, a card rotation occurred during this time, making 208 cards released prior to 2016 unusable in the standard play format. Such rotations, which happen yearly around April, are designed to prevent certain powerful cards and tactics from dominating the meta for too long, and allowing new content to be released without requiring to account for all previously released cards when testing for overpowered tactics. Two of the rotated cards, \textit{Emperor Thaurissan} and \textit{Aviana}, greatly improved the viability of \textit{Malygos Druid}. Thus the extinction of this archetype after $T=43$ in Figure \ref{fig:druidTL} could be anticipated. On the other hand, player experimentation also occurred due to the release of new content, with two new archetypes seeing significant play --- \textit{Ramp Druid} and \textit{Token Druid}. While \textit{Ramp Druid} lost popularity with players soon-after, \textit{Token Druid} continued as a popular \textit{Druid} archetype until a patch released after report $55$ \textit{nerfed} the card \textit{The Crystal Core}. This patch greatly affected the archetype \textit{Crystal Rogue}, causing it to fall out of the meta. \textit{Crystal Rogue} was one of \textit{Jade Druid}'s worst match-ups, in addition to being a very favourable one for \textit{Token Druid}. This flow-on effect led to the eventual disappearance of \textit{Token Druid}, and further cemented \textit{Jade Druid}'s popularity in the meta. 

\subsection{Archetype frequencies and win-rates}
In this work we label the set of all active archetypes in the \textit{Hearthstone} meta for a particular reporting period $T$ as 
\begin{equation}
    \mathcal{X}^T = \left\{ X^T_1, X^T_2, \dots, X^T_N \right\}, \;\; T \in \{ 30, 149\},
    \label{freqdef}
\end{equation}
where $N \equiv |\mathcal{X}^T|$. $T\in \{30,149\}$ corresponds to the \textit{Data Reaper Report} which was the source of the data \cite{VSDRR}. This spans approximately three years, being collected over the period 14 December 2016 to 27 December 2019.

\begin{figure}
    \centering
    \includegraphics[width=10cm]{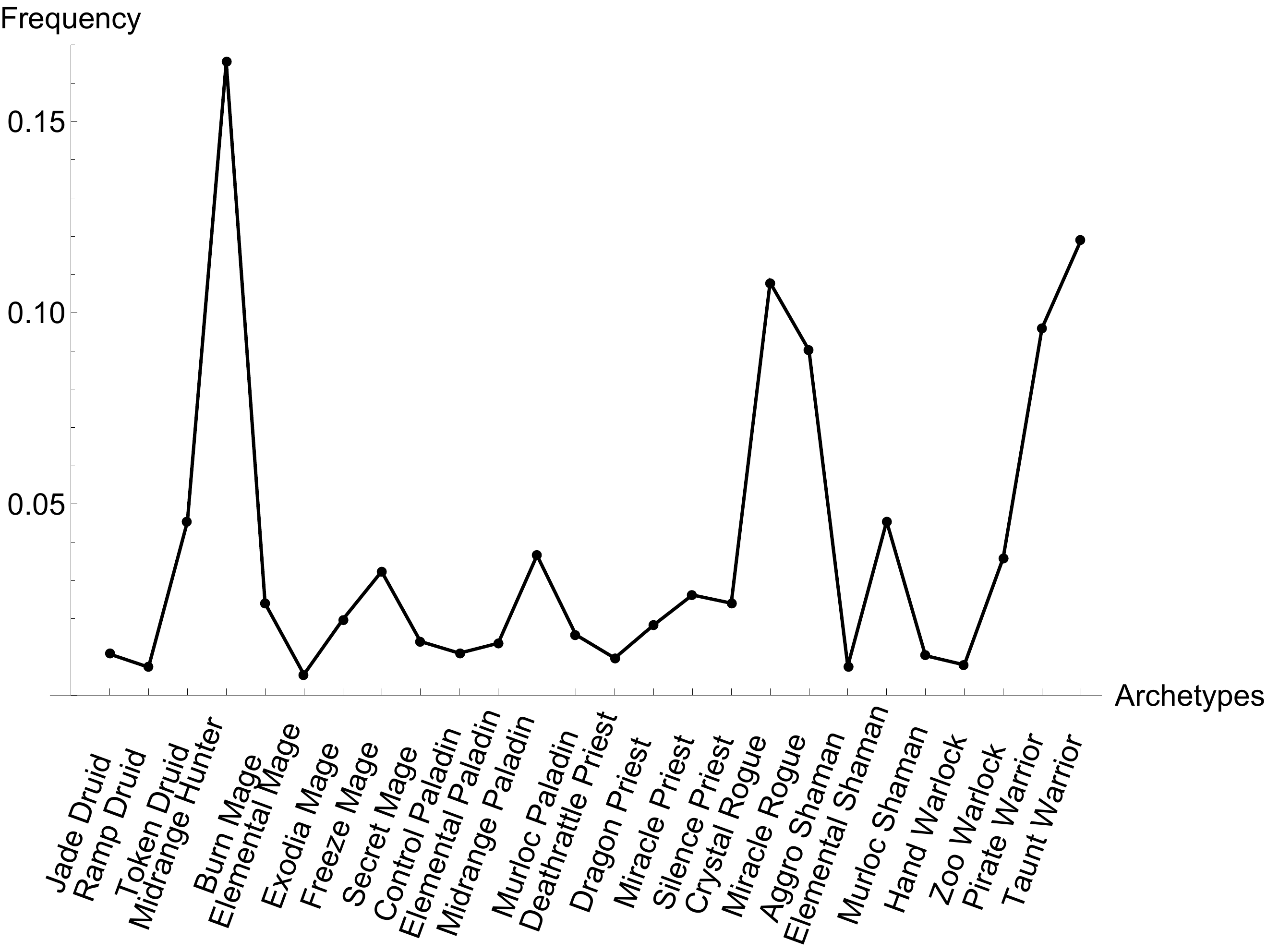}
    \caption{Archetype frequency data for $T=44$, collected over 10--18 April 2017.}
    \label{fig:P44}
\end{figure}

For each archetype $X^T_i$, the frequency that it was played in time period $T$ is labeled as $\mathcal{P}(X^T_i)$. For all $X^T_i \in \mathcal{X}^T$ the complete set of $\mathcal{P}$ forms a discrete probability distribution with the property 
\begin{equation}
    \sum^{|\mathcal{X}^T|}_{i=1}\mathcal{P}\left(X^T_i\right)=1 .
\end{equation}
Figure \ref{fig:P44} depicts the frequencies of active archetypes played over the period 10--18 April 2017, representing $T=44$ in Eq.(\ref{freqdef}). In this figure all character classes are represented in the 26 active archetypes, with the most frequently played archetype being \textit{Midrange Hunter}.

This work also considers the win-rates between archetypes, with $\mathcal{P}(W|P_{X^T_i},A_{X^T_j})$ denoting the conditional probability of \textit{winning}, given that \textit{player} ($P$) chose archetype $X_i$ and faced \textit{adversary} ($A$) using archetype $X_j$, at time period $T$. Win-rates and their transpose are equal to unity, leading to the identity
\begin{eqnarray}
\begin{split}
    &\mathcal{P}(W|P_{X^T_i},A_{X^T_j}) = 1 - \mathcal{P}(W|P_{X^T_j},A_{X^T_i}),\\
    \Rightarrow \;\; & \mathcal{P}(W|P_{X^T_j},A_{X^T_i}) = \mathcal{P}(L|P_{X^T_i},A_{X^T_j}),
\end{split}
\label{WLrel}
\end{eqnarray}
for $W=win$ and $L=lose$. Additionally mirror match-ups amongst the same archetype are equal to $0.5$, \textit{i.e.}
\begin{equation}
   \mathcal{P}(W|P_{X^T_i},A_{X^T_i}) = \mathcal{P}(L|P_{X^T_i},A_{X^T_i}) = 0.5.
\end{equation}

\subsection{Shannon information entropy}
\label{sec:shannon}

Shannon information entropy is measured via
\begin{equation}
    \mathcal{H}\left(\mathcal{X}^T \right) = - \sum^{|\mathcal{X}^T|}_{i=1} \mathcal{P}\left(X^T_i\right) \log_2  \mathcal{P}\left(X^T_i \right).
    \label{shannon}
\end{equation}
One of the main benefits of Shannon entropy is its ability in characterising the underlying complexity in the system \cite{Shannon49}. In general maximum entropy values are obtained by uniformly distributed probabilities --- \textit{i.e.} $\mathcal{P}(X_i) = 1/|\mathcal{X}| \;\; \forall \;\; X_i \in \mathcal{X}$. Thus Eq.(\ref{shannon}) offers insight into how evenly distributed the archetypes in the meta are over any given reporting period if compared the value of the theoretical maximum 
\begin{equation}
\mathcal{H}_{max}\left(\mathcal{X}^T \right) = \log_2  \left| \mathcal{X}^T \right|
\label{normalisedshan}
\end{equation}
which is the logarithm of the number of active deck archetypes in the meta for any given time period $T$.

\begin{figure}
    \centering
    \includegraphics[width=14.5cm]{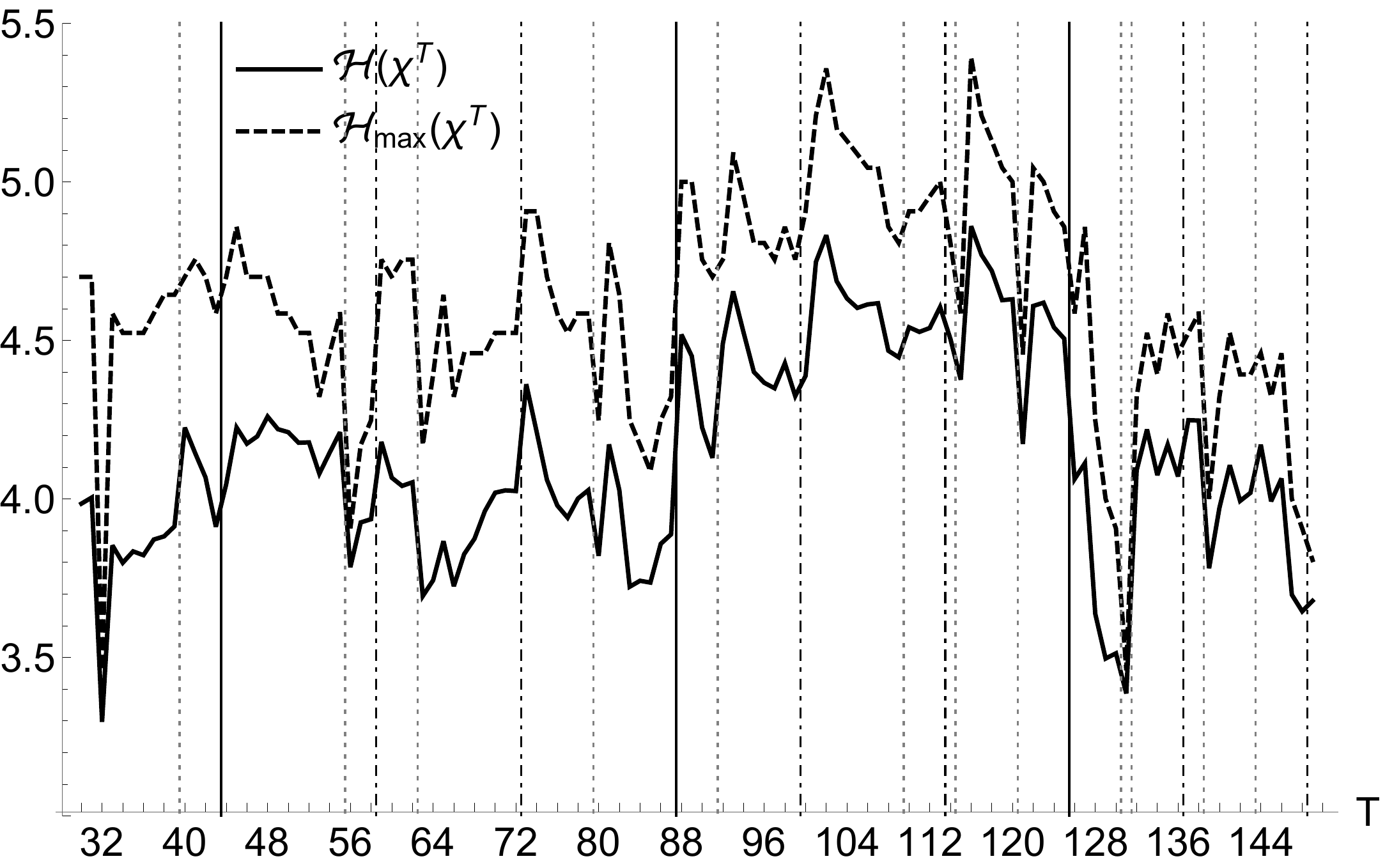}
    \caption{Graph of the Shannon information entropy (solid line) and its theoretical maximum (dashed line) --- defined in Eq.(\ref{shannon}) and (\ref{normalisedshan}) respectively --- derived from the frequencies that archetypes are played over each of the reporting periods. Refer to Figure \ref{fig:druidTL} for the specific meanings of vertical lines representing system changes.}
    \label{fig:shannonmaster}
\end{figure}

Figure \ref{fig:shannonmaster} depicts the Shannon entropy (solid line) and the corresponding maximum entropy (dashed line) derived from archetype frequencies for each of the reporting periods. Vertical lines represent changes to the system as explained in Section \ref{sec:timeline}. Both entropy values largely mirror each other over the entire time period which is expected. Noticeable increases and decreases occur immediately after a system-wide change has been introduced. The most common of these occurrences is a sharp increase, followed by a decrease in entropy until the next change occurs. This particular behaviour in entropy indicates a marked escalation in archetype experimentation immediately after changes to the \textit{Hearthstone} environment. System entropy decreases soon-after due to players understanding and settling on popular decks and tactics, which have emerged due to the changes. For the majority of Figure \ref{fig:shannonmaster}, this behaviour in entropy and the assumed player decision-making it stems from is repeated semi-consistently. 

There are instances in Figure \ref{fig:shannonmaster} where change led to a marked \textit{decrease} in entropy values. As previously mentioned in Section \ref{sec:timeline}, after $T=55$ in July 2017 a patch nerfed the card \textit{The Crystal Core}, which led to the extinction of \textit{Crystal Rogue} from the meta, along with other archetypes. Only the new \textit{Jade Rogue} emerged during this period, significantly decreasing the amount of active archetypes. A similar situation occurred after $T=62$ in September 2017 when a patch nerfed five cards. This decreased the frequency that \textit{Druid}-based archetypes were played. \textit{Mid-Token Druid} and \textit{Ramp Druid} (amongst others) were extinguished from the meta, with only the new \textit{Tempo Rogue} emerging during this period. Such events which corresponded to decreases in system entropy relate to relatively small changes. These changes targeted a few archetypes perceived to be overpowered. Nevertheless, marked decreases in entropy values did occur for the release of the \textit{Rastakhan's Rumble} (after $T=113$ in December 2018) and \textit{Rise of the Shadows} (after $T=125$ in April 2019) expansions which were major changes. Unlike other expansions their effect led to a reduction of the system's entropy as they both saw a drop in active archetypes present in the meta. For both of these cases, the large drops in entropy values were reversed due to the minor changes initiated after $T=114$ (December 2018) and $T=130$ (May 2019). Both of these patches greatly encouraged deck experimentation and saw the emergence of new archetypes.

\begin{figure}
    \centering
    \includegraphics[width=8.5cm]{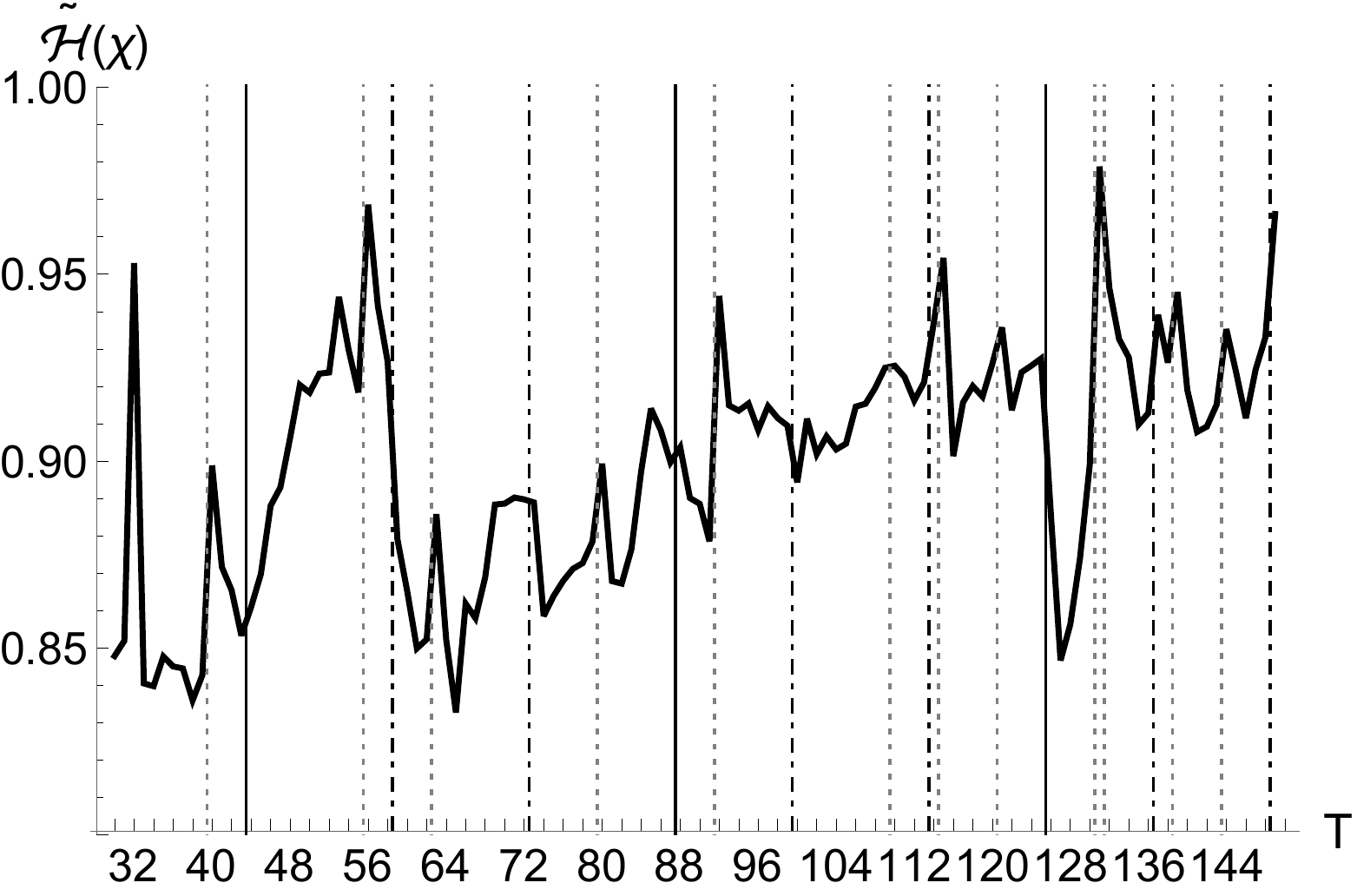}
    \caption{Graph of the normalised Shannon information entropy of the frequencies that archetypes are played for each reporting period. As with Figure \ref{fig:shannonmaster}, vertical lines represent system changes occurring between reporting periods.}
    \label{fig:shannonnormal}
\end{figure}

Figure \ref{fig:shannonnormal} depicts the normalised Shannon entropy, which is the Shannon entropy divided by the theoretical maximum value
\begin{equation}
\tilde{\mathcal{H}}(\mathcal{X}^T) =\mathcal{H}(\mathcal{X}^T)/ \mathcal{H}_{max}(\mathcal{X}^T).
\label{normalshannon}
\end{equation}
For any time period, Eq.(\ref{normalshannon}) varies between $(0,1)$, indicating how close the Shannon entropy is to the theoretical maximum. As with the values in Figure \ref{fig:shannonmaster}, this graph displays sharp variations when system changes are introduced. We conjecture that the evolution of the normalised Shannon entropy over time in Figure \ref{fig:shannonnormal} shows hallmark signs of a system at criticality \cite{hohenberg77,roli18}. The system responds and adapts to a rapidly changing environment, cycling through periods of relatively low and high normalised entropy values. High values indicate that all active deck archetypes are equally popular. If a deck's popularity indicates its likelihood to obtain victory, cases of high normalised entropy correspond to player choice offering little significance. This case correlates with all active archetypes being equally probable of obtaining victory, with a variety of equally viable tactics. Conversely, relatively low values indicate that only a small number of active archetypes are likely to be consistently victorious. Player choice would be biased towards those few archetypes, with experimentation kept to a minimum. System-wide changes occur to veer the \textit{Hearthstone} system away from these extreme situations \cite{Bursztein16}. The trajectory in Figure \ref{fig:shannonnormal} spends the majority of its time at intermediate values. This indicates a scenario where there are a range of viable archetypes, and player choice is not arbitrary as archetypes have varying strengths and weaknesses against each other. This scenario is reminiscent of Crutchfield and Young's \cite{crutch90} concept of the \textit{complexity spectrum} where a system displays the most complexity between its minimum and maximum normalised Shannon entropy values.

\section{Information storage and understanding player choice}
\label{sec:AIS}

\subsection{Local active information storage applied to archetype choices}

Following Eq.(\ref{LAISdef}), we construct the set of conditional probabilities $\mathcal{P}(X^T_i|X^{T-1}_{j_1},\dots, X^{T-K}_{j_K})$ which denote archetype $X^T_i$ being chosen, given that the archetypes $\{X^{T-1}_{j_1},\dots, X^{T-K}_{j_K}\}$ were played in the past. Importantly, the conditional probabilities possess the property of simulating understandable player behaviour. Additionally, the conditional probabilities must satisfy the consistency condition
\begin{equation}
    \mathcal{P}(X^{T}_i) = \left[\prod^{K}_{n=1} \sum^{|\mathcal{X}^{T-n}|}_{j_n=1}\mathcal{P}(X^{T-n}_{j_n})\right] \mathcal{P}(X^{T}_i|X^{T-1}_{j_1},\dots, X^{T-K}_{j_K}) 
    \label{consistmast}
\end{equation}
following Bayes' theorem \cite{MacKay03}. Using this notation, the LAIS associated with deck $X^T_i$, given past choices, is given via
\begin{equation}
    a_K(X^T_i|X^{T-1}_{j_1},\dots, X^{T-K}_{j_K} ) = \log_2 \frac{\mathcal{P}(X^T_i|X^{T-1}_{j_1},\dots,X^{T-K}_{j_K})}{\mathcal{P}(X^T_i)}.
    \label{LAISdeck}
\end{equation}
This work applies the convention that LAIS values are zero if $\mathcal{P}(X^T_i)=0$ --- \textit{i.e.} the archetype $X^T_i$ is not active for time period $T$. Additionally the AIS associated with each archetype, labelled $A^{(arch)}_K(X^T_i)$, is given as the LAIS expectation value over all past choices \begin{eqnarray}
\begin{split}
    A^{(arch)}_K(X^T_i) =\left[ \prod^K_{n-1} \sum^{|\mathcal{X}^{T-n}|}_{j_n=1} \mathcal{P}(X^{T-n}_{j_n}) \right]\mathcal{P}(X^T_i|X^{T-1}_{j_1},\dots,X^{T-K}_{j_K}) a_K(X^T_i|X^{T-1}_{j_1},\dots, X^{T-K}_{j_K} ).
    \end{split}
    \label{totaldeckais}
\end{eqnarray}
Archetype-AIS values in Eq.(\ref{totaldeckais}) are equivalent to the definition of \textit{agent rigidity} given in \cite{Cliff17}, used as a measure of agent predictability. Also used in this work is the total AIS for each time period, labelled $A^T_K$, which is the LAIS expectation value over all past and current archetype choices   
\begin{equation}
   A^T_K= \sum^{|\mathcal{X}^{T}|}_{i=1} A^{(arch)}_K(X^T_i).
    \label{totalais}
\end{equation}

\subsection{Weightings applied to construct conditional probabilities}
\label{ASSUM1}

This work is guided by the assumption that player decision-making is solely influenced by comparing past archetype frequencies and/or win-rates to adjust choices accordingly. Thus, if a player chooses archetype $X_j$ in time period $T-1$, then the probability that archetype $X_i$ is chosen in the next time period $T$ is weighted by a function of both archetypes in the previous time period. This is expressed mathematically via
\begin{equation}
    \mathcal{P}(X^{T}_{i}|X^{T-1}_{j}) = \epsilon^{(i)} \mathcal{P}(X^T_i) \mathcal{K}_{T-1}(i|j) , \;\; \;\;
    \textrm{where}\;\; \;\; \mathcal{K}_{T-1}(i|j) =f\left( X^{T-1}_{i}, X^{T-1}_j \right) .
    \label{assumpinformed}
\end{equation}
The weighting $\mathcal{K}$ compares deck-frequencies, and/or win-rates that players experienced against archetype $X_i$, given they played archetype $X_j$ in time period $T-1$. The function $f$ results in a larger weighting if archetype $X_i$ was played more frequently, and/or had a higher win-rate against $X_j$. Additionally, the coefficient $\epsilon^{(i)}$ ensures that the consistency condition given by Eq.(\ref{consistmast}) is satisfied. In order to test the assumption over multiple $K$-time periods the corresponding conditional probabilities are given via
\begin{equation}
     \mathcal{P}(X^{T}_i|X^{T-1}_{j_1},\dots, X^{T-K}_{j_K}) = \epsilon^{(i)} \mathcal{P}(X^T_i)
     \prod^K_{n=1}\mathcal{K}_{T-n}(i|j_n),
\label{informedassum}
\end{equation}
with the weighting factors $\mathcal{K}$ defined in Eq.(\ref{assumpinformed}). Additionally, by ensuring that the consistency condition given in Eq.(\ref{consistmast}) is adhered to, the coefficients $\epsilon^{(i)}$ are given as the following
\begin{equation}
    \epsilon^{(i)} =   \prod^{K}_{n=1}\left[ \sum^{|\mathcal{X}^{T-n}|}_{j_n=1} \mathcal{P}(X^{T-n}_{j_n}) \mathcal{K}_{T-n}(i|j_n)  \right]^{-1}.
\end{equation}
Thus, over a general number of $K$ time periods, the AIS associated with each archetype $X^T_i$, is
\begin{eqnarray}
\begin{split}
    A^{(arch)}_K(X^T_i) =\left[ \prod^K_{n-1} \sum^{|\mathcal{X}^{T-n}|}_{j_n=1} \mathcal{P}(X^{T-n}_{j_n}) \mathcal{K}_{T-n}(i|j_n) \right]\epsilon^{(i)} \mathcal{P}(X^T_i)
     \log_2  \epsilon^{(i)} \prod^K_{n-1} \mathcal{K}_{T-n}(i|j_n)  ,
     \end{split}
\end{eqnarray}
with the AIS for the entire time period $T$ given by Eq.(\ref{totalais}).

Eq.(\ref{functionalfr}) presents the exact forms of the weighting functions applied in Eq.(\ref{informedassum}) when considering past archetype frequencies $f_{FR}$:
\begin{eqnarray}
    \begin{split}
    f_{FR}\left( \Delta \mathcal{P} \right) = Char_{ij} \times \left\{ \begin{array}{l}
   e^{2 \textrm{sgn}\left(\Delta\mathcal{P}\right) \left|\Delta\mathcal{P}\right|} \\
    e^{ \textrm{sgn}\left(\Delta\mathcal{P} \right) \left|\frac{\Delta\mathcal{P}}{0.2}\right|^2}\\
    e^{2 \textrm{sgn}\left(\Delta\mathcal{P}\right) \left|\Delta\mathcal{P}\right|^{\frac{1}{2}}}\\
    e^{2 \textrm{sgn}\left(\Delta\mathcal{P}\right) \left|\Delta\mathcal{P}\right|^{\frac{1}{4}}}
    \end{array}\right. \;\;
    \textrm{where} \;\; \Delta \mathcal{P} =\left\{\begin{array}{c} \mathcal{P}\left( X^{T-1}_{i}\right) - \mathcal{P}\left( X^{T-1}_{j}\right)\\ i \ne j, \\
    \mathcal{P}\left( X^{T-1}_{i}\right) -  \bar{\mathcal{P}}(\mathcal{X}^{T-1}) \\
    i =j,
    \end{array}\right.
        \end{split}
    \label{functionalfr}
\end{eqnarray}
and 
\begin{equation}
\bar{\mathcal{P}}(\mathcal{X}^{T-1}) = \sum_{k=1}^{ \left|\mathcal{X}^{T-1}\right|}\frac{\mathcal{P}(X^{T-1}_k)}{ |\mathcal{X}^{T-1}|}
\end{equation}
is the mean value of the archetype frequencies played over time-period $T-1$. The term $Char_{ij}$ in Eq.(\ref{functionalfr}) is a multiplicative factor which checks the \textit{character class} of both archetypes, and returns a value greater than unity if the character classes are equal, or $1$ otherwise. As explained in Section \ref{deckarch} each of the character classes gain access to class specific cards which require a resource investment to both obtain, and learn how play effectively. Thus the term $Char_{ij}$ simulates the \textit{resource hurdle} and/or \textit{unwillingness} involved in changing character classes. This work sets $Char_{ii}=2$, assuming that players are doubly likely to choose a deck if it is the same character class as the deck they played in the previous time period.

Additionally, Eq.(\ref{functionalwr}) gives the forms of the weighting functions applied in Eq.(\ref{informedassum}) when considering past win-rates $f_{WR}$:
\begin{eqnarray}
    \begin{split}
    f_{WR}\left( \mathcal{P} \right) = \left\{ \begin{array}{l}
   e^{2 \textrm{sgn}\left(\mathcal{P}-0.5\right) \left|\mathcal{P}-0.5\right|} \\
    e^{2 \textrm{sgn}\left(\mathcal{P}-0.5\right) \left|\frac{\mathcal{P}-0.5}{0.5}\right|^2}\\
    e^{2 \textrm{sgn}\left(\mathcal{P}-0.5\right) \left|\mathcal{P}-0.5\right|^{\frac{1}{2}}}\\
    e^{2 \textrm{sgn}\left(\mathcal{P}-0.5\right) \left|\mathcal{P}-0.5\right|^{\frac{1}{4}}}
    \end{array}\right. \;\;
    \textrm{where} \;\; \mathcal{P} = \left\{\begin{array}{c} 
    \mathcal{P}\left(L \left| P_{X^{T-1}_{j}} \right.,A_{X^{T-1}_i}\right)\\
    i \ne j,\\
    \bar{\mathcal{P}}\left(L \left| P_{\mathcal{X}^{T-1}} \right.,A_{X^{T-1}_i}\right)\\
    i = j,
    \end{array} \right.
    \end{split}
    \label{functionalwr}
\end{eqnarray}
and
\begin{equation}\bar{\mathcal{P}}\left(L | P_{\mathcal{X}^{T-1}} ,A_{X^{T-1}_i}\right) = \sum_{k=1}^{ \left|\mathcal{X}^{T-1}\right|} \mathcal{P}(X^{T-1}_k) \mathcal{P}\left(L | P_{X^{T-1}_k},A_{X^{T-1}_i}\right)
\end{equation}
is the mean value of the win-rate for archetype $X_i$ over time-period $T-1$.

\begin{figure}
    \centering
    \includegraphics[width=15cm]{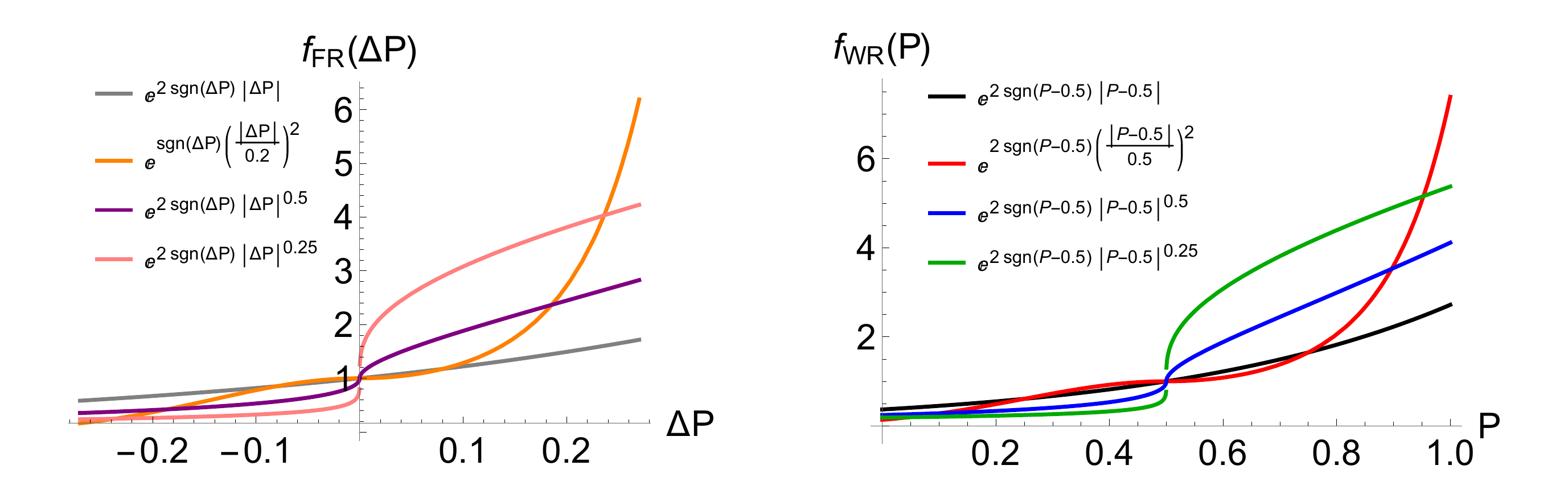}
    \caption{Plots of the specific forms of the functional responses applied to the weightings in Eq.(\ref{assumpinformed}). Left panel shows archetype frequencies $f_{FR}$ detailed in Eq.(\ref{functionalfr}), and the right panel shows win-rates $f_{WR}$ detailed in Eq.(\ref{functionalwr}).}
    \label{fig:functional1}
\end{figure}

Figure \ref{fig:functional1} depicts the functional responses applied to the weightings in Eq.(\ref{functionalfr}) (left panel) and Eq.(\ref{functionalwr}) (right panel). The grey and black trajectories in the left and right panels respectively display an almost-linear response. Thus for these functions the probability of choosing archetype $X_i$ rises linearly the more it was played (left panel) and the better it performed against $X_j$ in the previous time period (right panel). The remaining coloured trajectories present non-linear responses. In the left hand plot, the quadratic weighting in orange grows slowly initially, but then experiences the sharpest rise as $\Delta \mathcal{P} \rightarrow  0.27$. This range is chosen due to the largest frequency (occurring at $T=129$) being \textit{Lackey Rogue} played 26.5\% of the time. In contrast to this, the pink trajectory experiences its sharpest rise immediately after $\Delta \mathcal{P} = 0$, with a steady rise afterwards. An equivalent picture is presented with the four trajectories on the right hand panel of Figure \ref{fig:functional1}. If the win-rate is less than 50\% then the exponentials have a negative argument, leading to minimal weighting. If the win-rate is greater than 50\% the arguments are positive and the weightings grow non-linearly, for all but the black trajectories. Mirroring the left hand panel, the quadratic weighting in red grows slowly initially, experiencing the sharpest rise as $\mathcal{P} \rightarrow  1$. This range is chosen due to the largest win-rate (occurring at $T=123$) being \textit{Taunt Warrior} winning against \textit{Cube Rogue} 96.7\% of the time.

\subsection{Comparing archetype frequencies and win-rates individually}
\label{sec:frwr}

\begin{figure}
    \centering
    \includegraphics[width=14.5cm]{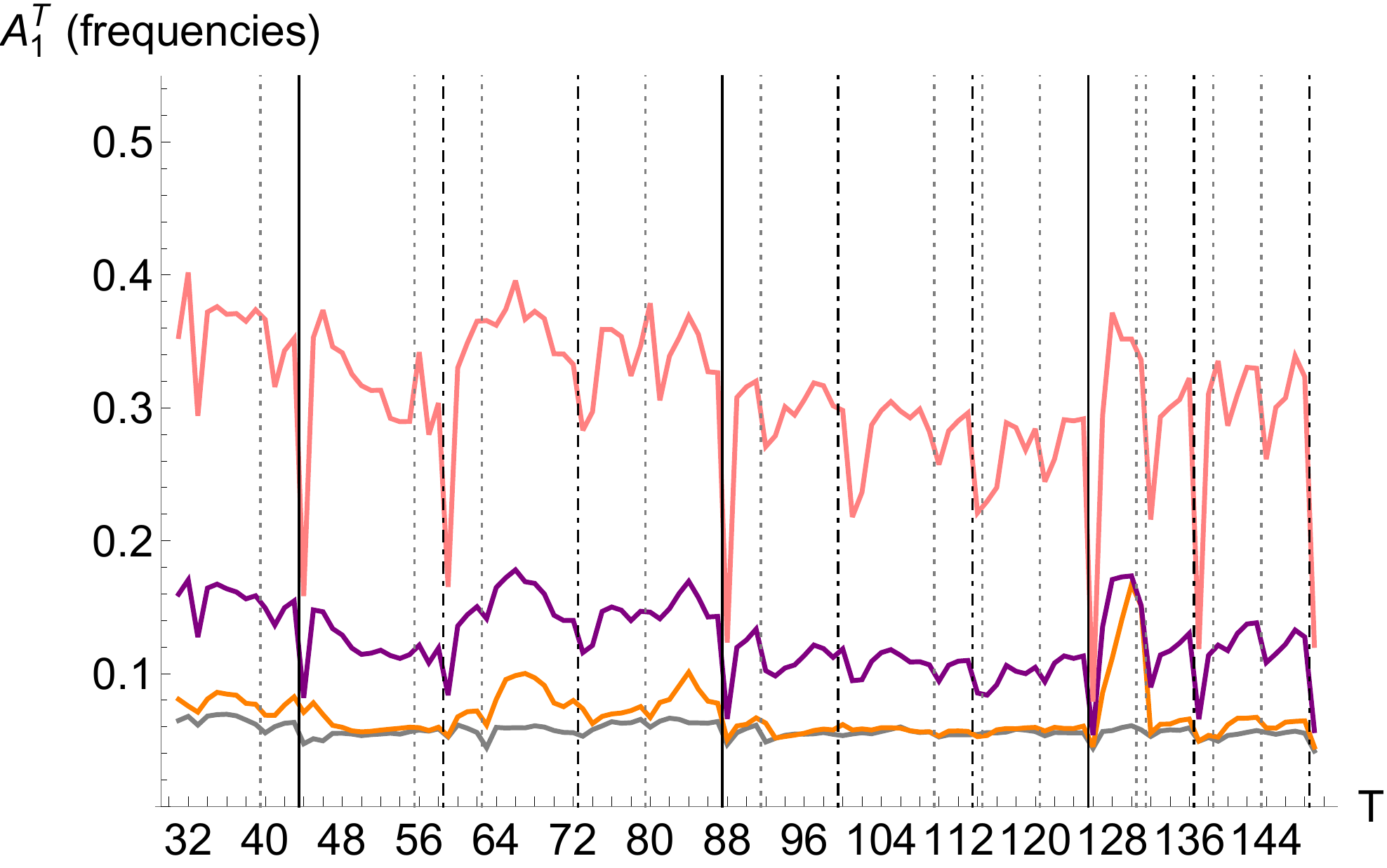}
    \caption{Plots of Eq.(\ref{totalais}) for $K=1$ for total AIS associated with each time period. Each trajectory was calculated assuming functional responses of the past archetype frequencies given in Eq.(\ref{functionalfr}). Note that the colours for each trajectory corresponds with the colours given to each functional response on the left panel of Figure \ref{fig:functional1}.}
    \label{fig:AISfr}
\end{figure}

Figure \ref{fig:AISfr} depicts the total AIS defined in Eq.(\ref{totalais}) for $K=1$. The functional responses of the past frequencies correspond to Eq.(\ref{functionalfr}), with the colours of each trajectory matching the colours given to each functional response in the left panel of Figure \ref{fig:functional1}. A major feature of Figure \ref{fig:AISfr} is the marked difference between the AIS values with different functional responses. The almost-linear (grey) and squared (orange) responses display similar AIS values, except at $T \in (124,132)$. The highest AIS values are obtained by the pink trajectory, whose functional response rises the sharpest as the difference between the frequencies becomes greater than 0, as witnessed in the left panel of Figure \ref{fig:functional1}. Thus, the functional response which rises the sharpest immediately after the archetype under consideration compares favourably best aligns with actual player behaviour.  

The AIS values in Figure \ref{fig:AISfr} also experience a significant decrease whenever they cross time periods where a major change is introduced into the system. Thus players base significantly less of their decision-making on past outcomes immediately after such change. This is illustrated by considering the five changes which happened between $T=39$ and $T=63$. At the end of February 2017 ($T=39$) a patch was released which nerfed the cards \textit{Small Time Buccaneer} and \textit{Spirit Claws} in order to break the dominance of \textit{Aggro Shaman} in the meta. This patch, as well as the patches released after $T=55$ (July 2017) and $T=62$ (September 2017), only affected a handful of cards and were intended to affect a small number of archetypes in the meta. Though such small changes substantially changed the meta and its corresponding Shannon entropy (as shown in Figures \ref{fig:P44} and \ref{fig:shannonmaster}), these changes did not substantially change the decision-making players employed to choose archetypes, having very little effect on the AIS values. Interestingly, AIS values actually increased immediately after $T=55$, meaning this change actually \textit{reinforced} past decision-making. This is in stark contrast to AIS values occurring immediately after $T=43$ (April 2017) and $T=58$ (August 2017), with the release of the \textit{Journey to Un'goro} and \textit{Knights of the Frozen Throne} expansions, respectively. In fact, all of the major decreases in AIS values in Figure \ref{fig:AISfr} occurred immediately after significant changes were introduced.

\begin{figure}
    \centering
    \includegraphics[width=14.5cm]{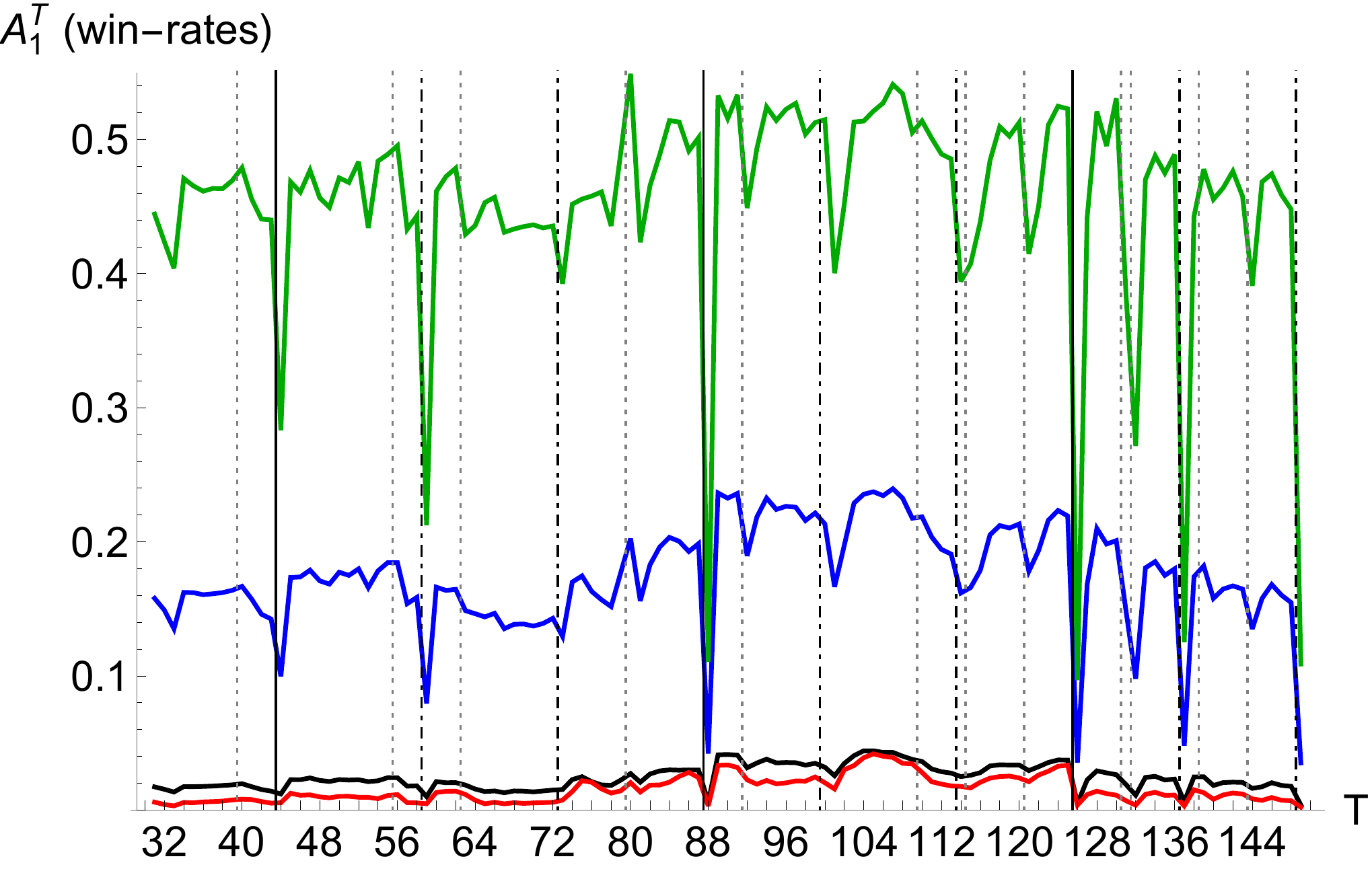}
    \caption{Plots of Eq.(\ref{totalais}) for $K=1$ for total AIS associated with each time period. Each trajectory was calculated assuming functional responses of the past win-rates given in Eq.(\ref{functionalwr}). Note that the colours for each trajectory corresponds to the colours given to each functional response on the right panel of Figure \ref{fig:functional1}.}
    \label{fig:AISwr}
\end{figure}

Figure \ref{fig:AISwr} presents the total AIS defined in Eq.(\ref{totalais}) for $K=1$. Functional responses of the past win-rates are given in Eq.(\ref{functionalwr}). The colours of each trajectory in Figure \ref{fig:AISwr} matches the colours in the right panel of Figure \ref{fig:functional1}. Figure \ref{fig:AISwr} displays many of the same features already discussed in Figure \ref{fig:AISfr}. These include notable decreases in AIS values immediately after significant change is introduced, and subdued responses (or slight increases) for minor change. Also, the highest AIS values are obtained by the green trajectory, which has a similar functional response to the pink trajectory which obtains the highest AIS values in Figure \ref{fig:AISfr}. 

\subsection{Combining archetype frequencies and win-rates}
\label{sec:comb}
\begin{figure}
    \centering
    \includegraphics[width=14.5cm]{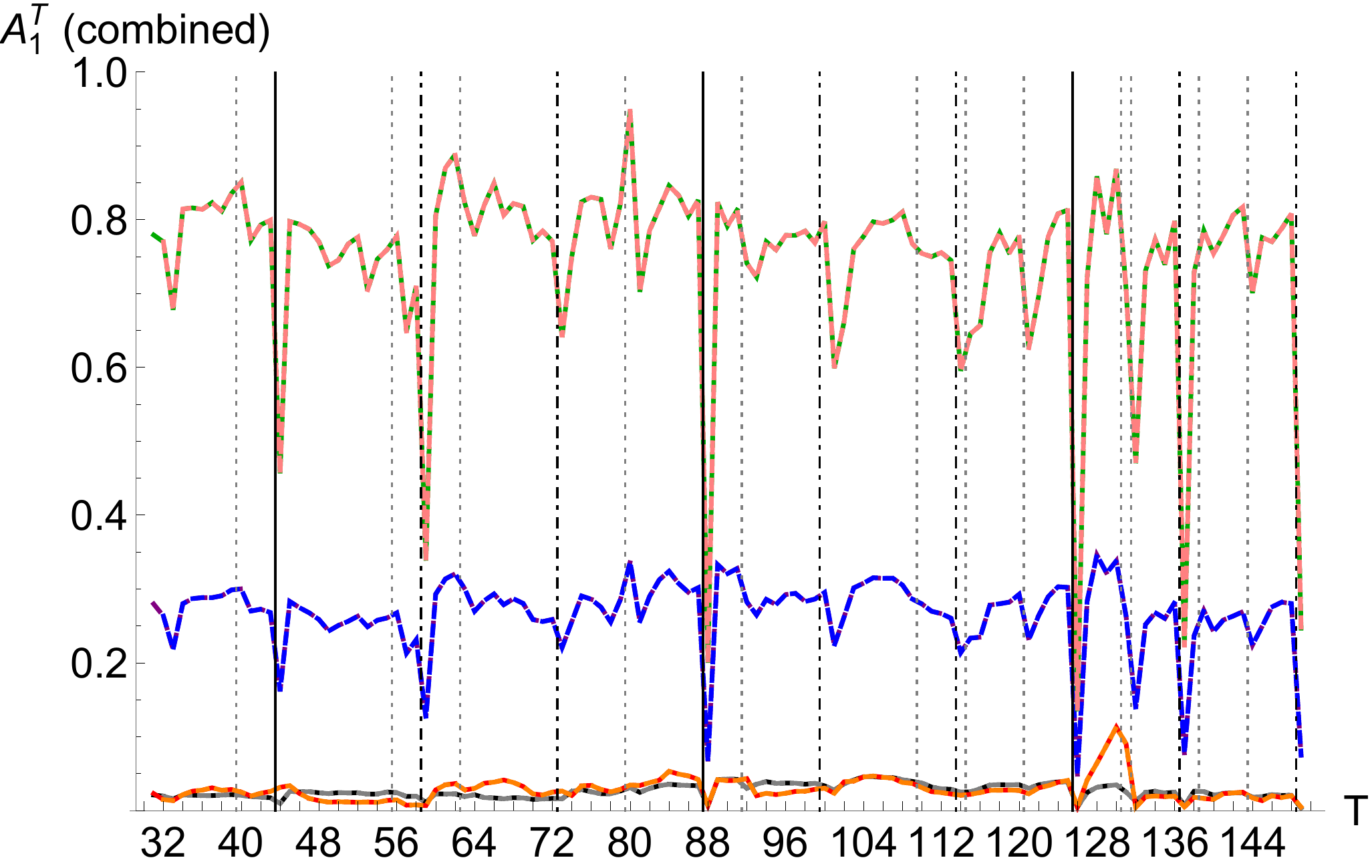}
    \caption{Plots of Eq.(\ref{totalais}) for $K=1$ for total AIS associated with each time period. The functional responses, given in Eq.(\ref{functionaltotal}), compare both archetype frequencies and win-rates. Each trajectory is denoted by two colours which signify which of the functional responses were combined on the left hand panel (archetype frequencies) and right hand panel (win-rates) of Figure \ref{fig:functional1}.}
    \label{fig:AIStot}
\end{figure}

Figure \ref{fig:AIStot} presents the AIS values which combine the archetype frequency and win-rate functional responses given by
\begin{equation}
    \mathcal{K}_{T-1}(i|j) = \underbrace{f_{FR} (\Delta \mathcal{P})}_{\textrm{Eq.(\ref{functionalfr})}} \times \underbrace{f_{WR} (\mathcal{P})}_{\textrm{Eq.(\ref{functionalwr})}}.
    \label{functionaltotal}
\end{equation}
Each trajectory in Figure \ref{fig:AIStot} is composed of two colours. These signify which of the functional responses were combined on the left hand panel (archetype frequencies) and right hand panel (win-rates) of Figure \ref{fig:functional1}. The trajectories in Figure \ref{fig:AIStot} generally display similar features to those seen in Figures \ref{fig:AISfr} and \ref{fig:AISwr}. The main difference however is the marked increase in AIS values in Figure \ref{fig:AIStot}, approximately doubling the values seen in Figures \ref{fig:AISfr} and \ref{fig:AISwr}.

\begin{figure}
    \centering
    \includegraphics[width=8.5cm]{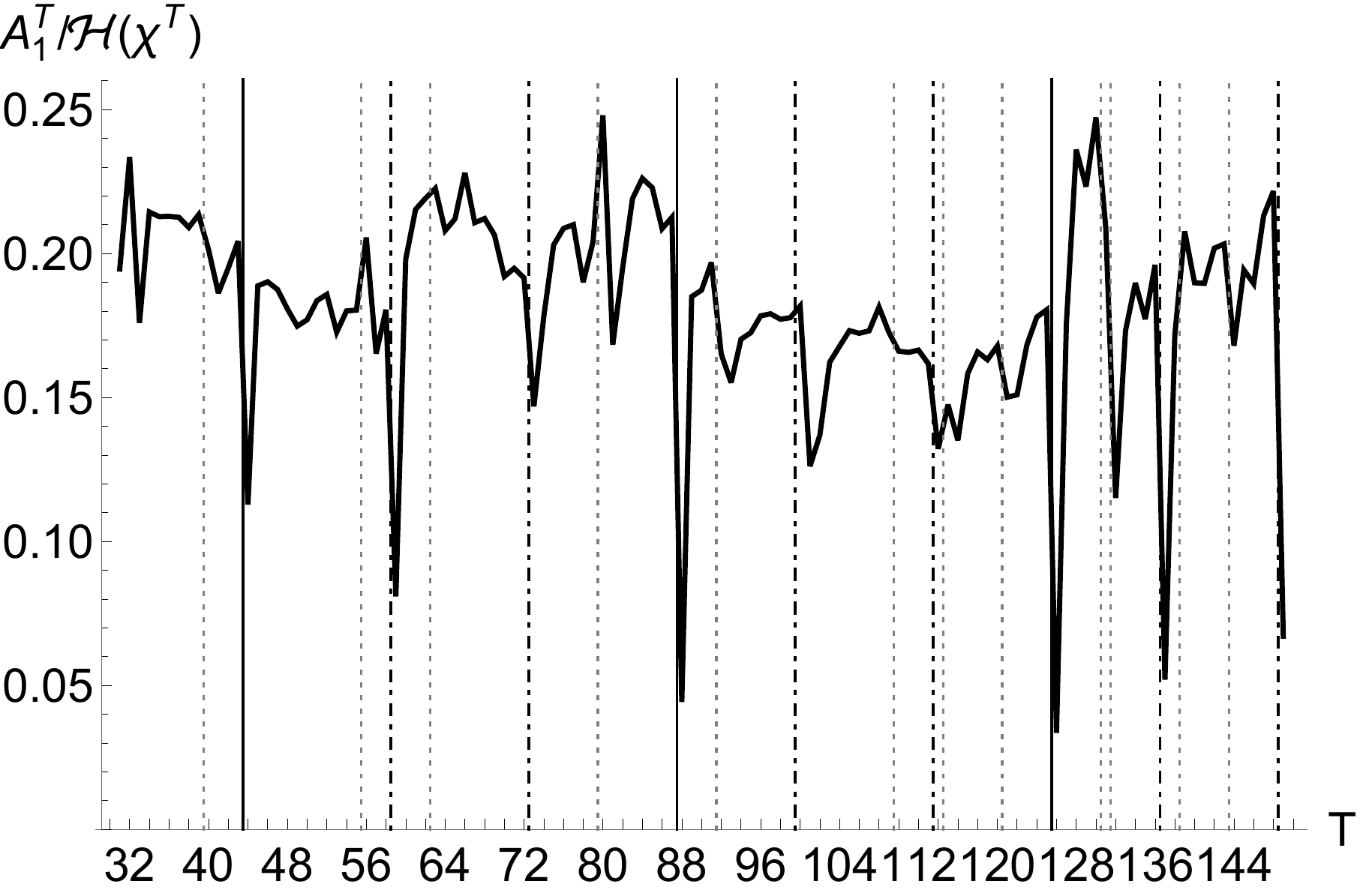}
    \caption{Plot of the percentage of uncertainty (or surprise) within the \textit{Hearthstone} meta that is explained by considering past archetype frequencies and win-rates via the functional response given in Eq.(\ref{functionaltotal}).}
    \label{fig:explain}
\end{figure}
Recalling the duality relation between Shannon entropy, AIS and the entropy rate in Eq.(\ref{duality}), we can compare the AIS values in Figure \ref{fig:AIStot} with the Shannon entropy in Figure \ref{fig:shannonmaster}. Doing so enables us to appreciate how much of the current state of the \textit{Hearthstone} meta is captured by the  functional responses we have used to simulate player decision-making. Figure \ref{fig:explain} plots the exact value of this explain-ability ($A^T_1 / \mathcal{H}(\mathcal{X}^T)$) per time period, for the highest $A^T_1$ values taken from the green-pink trajectory in Figure \ref{fig:AIStot}. The value of $A^T_1 / \mathcal{H}(\mathcal{X}^T)$ varies between $[0,1]$ for any system. A value close to zero signifies that the assumptions used to construct the AIS reveal very little about the current state of the system. Likewise, a value close to unity signifies that the assumptions applied to construct the AIS offers a near-to-complete explanation of the current state of the system. The assumptions in Eq.(\ref{functionaltotal}) are designed to simulate understandable player behaviour, with more popular and better performing archetypes having greater probability of being chosen. For most time periods, Figure \ref{fig:explain} shows that this simple principle explains approximately 20\% of the \textit{Hearthstone} meta. Though our assumptions do not take into account the nuances of player motivations when faced with deck construction, the fact that AIS values drop so dramatically immediately after large changes validates our assumptions. As discussed in Sections \ref{sec:timeline} and \ref{sec:shannon}, deck construction and tactics experimentation generally increases immediately after such changes, leading to the emergence of new archetypes, and the corresponding Shannon entropy. Hence, during these periods it would be incorrect to assume that relying on past results to inform current decisions would lead to good outcomes. The dramatic decreases in AIS values during these periods in Figures \ref{fig:AISfr}, \ref{fig:AISwr} and \ref{fig:AIStot} validates these assertions.

\subsection{Active information storage of archetypes}
\begin{figure}
    \centering
    \includegraphics[width=14.5cm]{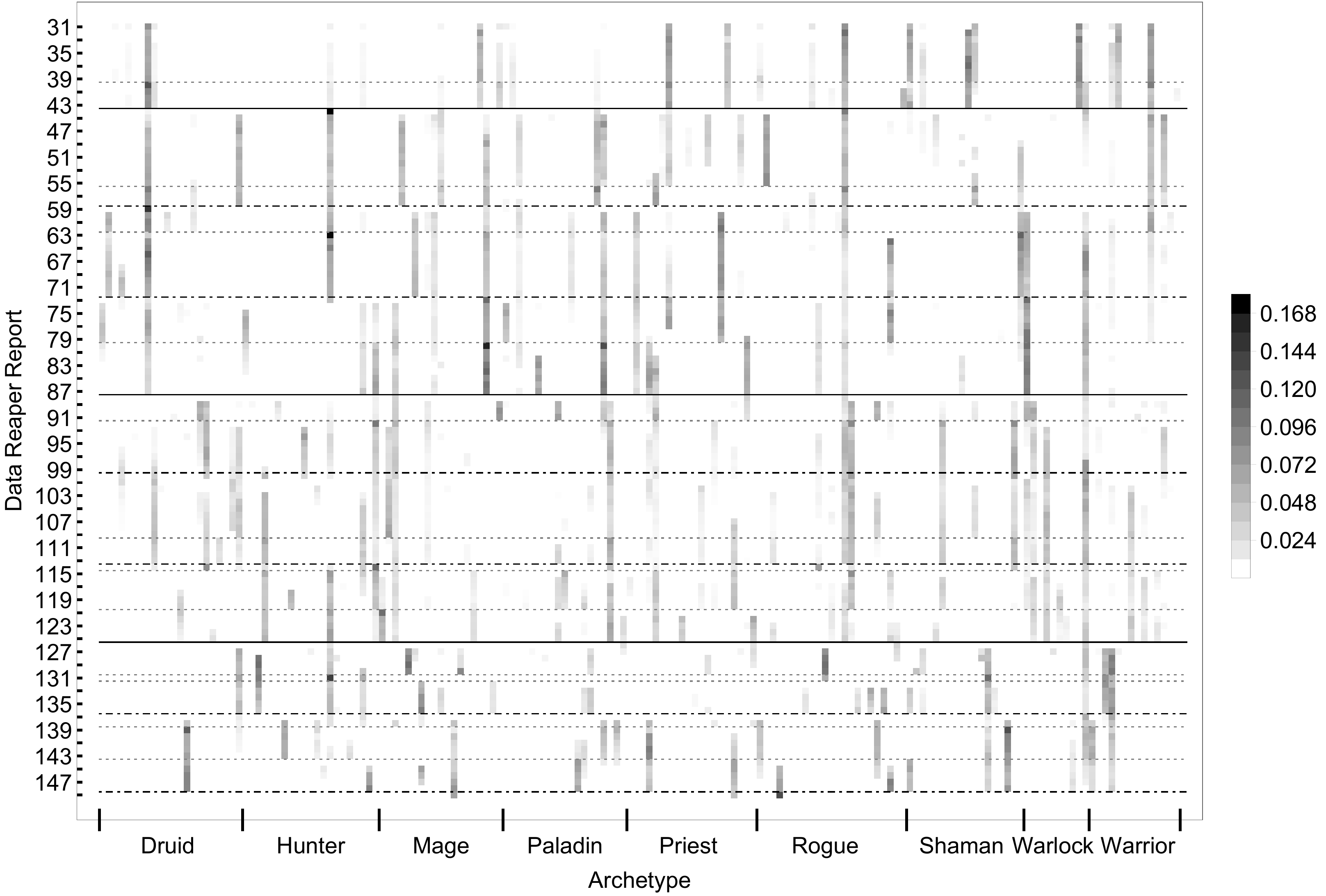}
    \caption{Plot of Eq.(\ref{totaldeckais}), giving the archetype-AIS values per time period which were used to generate the pink-green total AIS values in Figure \ref{fig:AIStot}.}
    \label{fig:deckAIS}
\end{figure}

Figure \ref{fig:deckAIS} provides a heat-plot of the archetype AIS values per time period --- $ A^{(arch)}_1(X^T_i)$ via Eq.(\ref{totaldeckais}) --- which were used to generate the largest combined AIS values (pink-green) in Figure \ref{fig:AIStot}. The horizontal axis of Figure \ref{fig:deckAIS} indicates each of the $166$ deck archetypes considered in this study, which are ordered alphabetically within each of the nine character classes. The vertical axis indicates the \textit{data reaper report} for the archetype AIS values. Horizontal lines in Figure \ref{fig:deckAIS} indicate changes occurring between reporting periods, as per the convention detailed in Section \ref{sec:timeline}. White regions signify archetypes that did not contribute AIS values for that particular time period. Non-zero AIS values signify that the frequency of play of these archetypes at time $T$ correlates with the state of the meta at time $T-1$. The darker the colour, the more pronounced the correlation. 

Figure \ref{fig:deckAIS} reveals in greater detail the impact that change has on the various archetypes, useful when comparing to the global picture given in Figures \ref{fig:AISfr}--\ref{fig:explain}. Periods experiencing small changes generally have minimal effect on the AIS values of the majority of the active archetypes in the meta. Contrast to this, time periods experiencing large changes generally display disruptive effects, with a sizeable proportion of archetypes receiving zero AIS values, and a noticeable change in the AIS values of the remaining active archetypes. Focusing again on the major change occurring immediately after $T=43$, Figure \ref{fig:AIStot} displayed a sizable drop in total AIS values due to the \textit{Journey to Un'goro} expansion and card rotation. The impacts of these changes are made clearer in Figure \ref{fig:deckAIS}, with a number of the active archetypes receiving no deck-AIS values past $T=43$. Nevertheless, archetypes which survive to $T=44$ actually obtain a sizable increase in AIS values, such as \textit{Midrange Hunter} and to a lesser extent \textit{Miracle Rogue}. Due to these archetypes surviving the change and performing relatively well in the meta, we interpret the relatively large increase in AIS values as these archetypes offering players a means to reinforce their previous decision-making during a disruptive period.

\section{Discussion and future work}
\label{conc}
In this work we applied a number of information-theoretic measures to characterise and understand three years of game data of the online CCG \textit{Hearthstone}. Producing the system's Shannon entropy using the frequencies that deck archetypes are played provided a unique and useful characterisation of \textit{Hearthstone's} meta. One striking trend which manifested across the majority of the time-period was that most of the variability in the entropy appeared immediately after a system-wide change had occurred. Sharp increases in entropy values, usually followed by decreases immediately after, implied a marked escalation in deck construction experimentation after change had been initiated. Entropy decreasing soon-after can then be understood as players understanding and exploiting the strong decks and tactics which emerged due to these changes.

Additionally, by constructing conditional probabilities that particular archetypes were chosen in the current time-period based on the past state of the system, we examined the information storage exhibited in \textit{Hearthstone}'s meta. Importantly, the weightings used to construct the conditional probabilities simulated simple player decision-making. An undeniable feature which emerged from the resulting AIS values were the significant decreases experienced during periods of major change, implying that players base significantly less of their decision-making on past outcomes during disruptive periods. Furthermore, small system changes did not significantly change the underlying decision-making players employed in their archetype choices. In some instances AIS values actually increased, implying that such changes effectively reinforced past player decision-making. 



There are a number of avenues to further this work, both for CCGs and wider application areas.  Similar to \cite{Cliff17}, it may be possible to combine exploration of information transfer entropy and AIS in an attempt to establish if the \textit{Hearthstone} system in Figure \ref{fig:landscape} displays the primitives (storage and communications) of a universal computer \cite{Feldman08}. One could also consider \textit{Fisher entropy} \cite{Prokopenko11} in an attempt to uncover control parameter(s) which influences CCG-system criticality. An additional generalisation would include trying to algorithmically-optimise AIS values by producing weights to replace the mathematical functions --- Eqs.(\ref{functionalfr}) and (\ref{functionalwr}) --- used in this work. This optimisation would come with the challenge of interpreting the results through the lens of understandable player behaviour \cite{Hoffman18}. It would also be meaningful to consider the impact of constructing conditional probabilities based on archetype choices beyond $T-1$ --- \textit{i.e.} $K > 1$ in Eq.(\ref{LAISdeck}). As mentioned in \cite{Wibral2014} the choice must be made carefully, since using too many past states can result in overestimation of the AIS value. Another novel avenue would include an information-theoretic extension of the algorithmic deck construction work of Fontaine \textit{et al.} \cite{Fontaine19} by including generalised entropies similar to those considered in Prokopenko \textit{et al.} \cite{Prokopenko06} to maximise synchronisation/coordination in artificial systems with the intent of \textit{information-driven evolutionary design}.

Applying similar methods to other game-related application areas, we posit that it would be possible to gain appreciation of the evolution of other games with online systems similar to Figure \ref{fig:landscape}. An equivalent analysis of the real-time-strategy game \textit{Starcraft II}, with its mix of human players and AI \cite{Arulkumaran19}, may offer non-trivial insights on the impacts of AI interacting with wider society. Finally we hope that the information-theoretic results obtained about the nature of decision-making behaviour in epochs of system-wide change will be used to examine relevant data sets stemming from wider society. Such applications include: understanding the economical impacts of shifts in the international political landscape \cite{Tzachi03}; and awareness of the changing nature of population behaviours \cite{Dexter05}.

\section*{Acknowledgements}
This work was supported by a Research Fellowship under Defence Science and Technology Group's Modelling Complex Warfighting strategic research initiative. The authors are grateful to the \textit{Vicious Syndicate} team for agreeing to share their data set for research purposes. We additionally acknowledge Ivan Garanovich, Scott Wheeler, Keeley McKinlay, Carlos Kuhn, Alexander Kalloniatis, Sean Franco and Daniela Schlesier for fruitful discussions, and Mikhail Prokopenko for helpful feedback on an early draft of this work.

\end{document}